\documentstyle[twoside,fleqn,espcrc2]{article}
\input{psfig}
\def\bold#1{\setbox0=\hbox{$#1$}%
     \kern-.025em\copy0\kern-\wd0
     \kern.05em\copy0\kern-\wd0
     \kern-.025em\raise.0433em\box0 }
\def\slash#1{\setbox0=\hbox{$#1$}#1\hskip-\wd0\dimen0=5pt\advance
       \dimen0 by-\ht0\advance\dimen0 by\dp0\lower0.5\dimen0\hbox
         to\wd0{\hss\sl/\/\hss}}

\newcommand{\be}{\begin{equation}}
\newcommand{\ee}{\end{equation}}
\newcommand{\bea}{\begin{eqnarray}}
\newcommand{\eea}{\end{eqnarray}}

\newcommand{\AmS}{{\protect\the\textfont2
  A\kern-.1667em\lower.5ex\hbox{M}\kern-.125emS}}

\title{B decays to excited charm mesons\\}

\author{P. Colangelo$^a$\thanks{Speaker at the Conference}, 
F. De Fazio$^a$ and N. Paver$^b$\\
\vskip 0.2cm
$^a$ Istituto Nazionale di Fisica Nucleare - Sezione di Bari, 
via Amendola n.173, 70126 Bari, Italy\\ 
\vskip 0.2cm
$^b$ Dipartimento di Fisica Teorica dell'Universit\'a di Trieste and
Istituto Nazionale di Fisica Nucleare - Sezione di Trieste, 
Strada Costiera 11, 34014 Trieste, Italy\\
}
       
\begin{document}

\begin{abstract}
We review several aspects of the phenomenology of P-wave $\bar q Q$ mesons:
mass splittings, effective strong couplings and leptonic constants.
We also describe a QCD sum rule determination to order $\alpha_s$
of the form factor $\tau_{1/2}(y)$ 
governing the semileptonic $B$ decays to the charm doublet with
$J^P=(0^+_{1/2},1^+_{1/2})$. 

\end{abstract}

\maketitle

\section{Spectroscopy and strong decays of mesons containing one heavy quark}

The spectroscopy of hadrons containing one heavy quark Q,
in the infinite $m_Q$ limit, is simplified 
by the decoupling of the heavy quark spin $\vec s_Q$
from the angular momentum of the light degrees of freedom
(quarks and gluons) $\vec s_\ell= \vec J - \vec s_Q$.
In this heavy quark theory (HQET) the states are classified 
by both $\vec J$ and  $\vec s_\ell$, and the
hadrons corresponding to the same $s_\ell$ belong to degenerate doublets. 
In the case of mesons,
the low-lying states $s_\ell^P={1\over 2}^-$ correspond to the 
pseudoscalar $0^-$ and vector $1^-$ mesons
($B,B^*$; $D,D^*$), the $s$-wave states of the constituent quark model.
The four states corresponding to orbital angular momentum $L=1$ 
can be classified in two doublets: 
$J^P=(0^+_{1/2},1^+_{1/2})$ and
$J^P=(1^+_{3/2},2^+_{3/2})$, which differ by the values 
$s_\ell^P={1\over 2}^+$ and $s_\ell^P={3 \over 2}^+$, respectively
\cite{review}.

The charmed $2^+_{3/2}$ state, denoted as $D_2^*(2460)$, 
has been experimentally observed: 
$m_{D_2^*}=2458.9\pm 2.0$ MeV, $\Gamma_{D_2^*}=23\pm 5$ MeV and
$m_{D_2^*}=2459\pm 4$ MeV, $\Gamma_{D_2^*}=25^{+8}_{-7}$ MeV for the neutral 
and charged states, respectively. 
The state $1^+_{3/2}$ can be identified with $D_1(2420)$, with
$m_{D_1}=2422.2\pm 1.8$ MeV and 
$\Gamma_{D_1}=18.9^{+4.6}_{-3.5}$  MeV,
neglecting a possible $1^+_{1/2}$ component allowed by
 the finite value of the 
charm quark mass. Experimental evidence of beauty
$s_\ell^P={3\over 2}^+$ states has also been reported \cite{pdg}.
The narrow width of both the states $2^+_{3/2}$ and $1^+_{3/2}$ 
is due to their $d-$wave suppressed strong transitions, governed
by strong coupling constants that
can be determined using  experimental information \cite{falk92}.

The not yet observed charm doublet $J^P=(0^+_{1/2},1^+_{1/2})$
($D_0, D^*_1$) has $s$-wave strong transitions.
Analyses of 
the couplings governing the two-body hadronic decays to pions
can be done by QCD sum rules, in a theoretical framework
where both the QCD heavy flavour-spin symmetry and  chiral
symmetry are implemented \cite{review}. Three effective couplings, in 
the infinite $m_Q$ limit, are relevant: $g$ 
(for the vertex $D^* D \pi$), $g^\prime$ (for $D_1^* D_0 \pi$)
and $h$ (for $D_1^* D^* \pi$). The numerical results
$g=0.2-0.4$, $g^\prime=0.07-0.13$ and $|h|=0.4-0.8$
 allow to predict 
$\Gamma(D_0^0 \to D^+ \pi^-) \simeq 180$ MeV and
$\Gamma(D_1^{*0} \to D^{*+} \pi^-) \simeq 165$ MeV, and
the mixing  angle $\alpha\simeq 16^0$ between $D^*_1$ and  $D_1$.
The predictions for the beauty sector are
$\Gamma(B_0 \to B^+ \pi^-) \simeq \Gamma(B_1^{*0} \to B^{*+} \pi^-) 
\simeq 360$ MeV \cite{colangelo95}.
\section{Semileptonic B decays to excited states and universal form factors}
The decay matrix elements governing 
$B \to (D_0, D_1^*) \ell \bar \nu $  
can be parameterized by six form factors which allow to compute
the  physical observables such as, e.g., the spectrum of the
momentum transfer to the lepton
 pair and of the charged lepton energy:
\bea
{ <D_0(v')|{\bar c}\gamma_\mu \gamma_5 b|B(v)> 
\over {\sqrt {m_B m_{D_0}}}  } &=& \nonumber\\
g_+(y) (v+v^\prime)_\mu +
 g_-(y) (v-v^\prime)_\mu \;\;\;&,& \nonumber 
\eea
\bea
{<D^*_1(v',\epsilon)|{\bar c}\gamma_\mu(1-\gamma_5)b|B(v)>
\over {\sqrt {m_B m_{D^*_1}}} } &=& \nonumber \\
g_{V_1}(y) \epsilon^*_\mu +\epsilon^* \cdot v \; [g_{V_2}(y) v_\mu+
g_{V_3}(y) v^\prime_\mu]&-& \nonumber \\
i \; g_A(y)
\epsilon_{\mu \alpha\beta \gamma} \epsilon^{*\alpha} v^\beta v^{\prime \gamma}
\;\;\; &;&  \nonumber \label{full}
\eea
$v$ and $v^\prime$ are the meson four-velocities ($y=v \cdot v^\prime$). 
As in the case of $B\to (D, D^*) \ell \bar\nu$ transitions
\cite{review},
in the limit $m_{b,c}\to \infty$ 
the form factors $g_i(y)$ can be related 
to a single universal function $\tau_{1/2}(y)$ 
through short-distance coefficients, perturbatively calculable, 
which depend on the heavy quark masses $m_{b,c}$,
on $y$ and on a renormalization mass-scale $\mu$ \cite{isgur91}.
Such relations, at the next-to-leading 
logarithmic accuracy in $\alpha_s$, are reported in 
\cite{colangelo98}.
A $\mu$ independent quantity $\tau_{1/2}^{ren}(y)$ 
can also be defined.

Similarly, the eight form factors relevant to the decays
$B \to D_1 \ell \bar \nu$ and
$B \to D^*_2 \ell \bar \nu$ can be related to the universal
function $\tau_{3/2}(y)$ \cite{isgur91}. 
The main difference with respect to the universal 
Isgur-Wise $B \to D,D^*$ form factor  $\xi(y)$, constrained by the heavy 
quark symmetry at the zero recoil point $\xi(1)=1$,
is that one cannot invoke symmetry arguments to predict the 
normalization of neither $\tau_{1/2}(y)$ nor $\tau_{3/2}(y)$.

The QCD sum rule determination
of $\tau_{1/2}$ including
${\cal O}(\alpha_s)$ corrections 
has been recently considered \cite{colangelo98}
(analogous corrections for the Isgur-Wise form factor $\xi$
were previoulsy determined in \cite{neubert93}).
Notice that this a genuinely field theoretical approach in the based on QCD,
The relevant three-point correlator is 
\bea
\Pi_\mu(\omega, \omega^\prime, y) &=& 
i^2 \int dx \; dz e^{i(k' x-k z)} \nonumber \\
&<&0|T[ J_s^{v'}(x), {\tilde A}_\mu(0), J_5^{v}(z)^\dagger]|0> \;\;, 
\label{threep}\nonumber
\eea
with
${\tilde A}_\mu= \bar h_{Q^\prime}^{v'} \gamma_\mu \gamma_5 h_Q^v$ the 
$b \to c$ weak axial current in HQET,
$J_s^{v'}=\bar q h_{Q^\prime}^{v\prime}$ and 
$J_5^{v}=\bar q i \gamma_5 h_Q^v$  
local  interpolating currents for the $D_0$ and $B$.
The residual momenta $k, k^\prime$, obtained by
the expansion of the heavy meson momenta in terms of the 
four-velocities:
$P=m_Q v+ k$, $P^\prime=m_{Q^\prime} v^\prime+ k^\prime$, 
are ${\cal O}(\Lambda_{QCD})$, and remain finite in 
the heavy quark limit. 

The  analyticity of 
$\Pi(\omega, \omega^\prime, y)$ in the variables
$\omega=2 v\cdot k$ and $\omega^\prime=2 v^\prime\cdot k^\prime$ at fixed $y$,
allows to express the correlator in terms of 
hadronic contributions,
from poles at positive real
values of $\omega$ and $\omega^\prime$ and from a continuum of 
states. 
The lowest-lying contribution corresponding to the 
$B$ and $D_0$ poles, introduces 
the form factor $\tau_{1/2}$ through the relation:
\begin{equation}
\Pi_{pole}(\omega, \omega^\prime, y) = 
{-2 \tau_{1/2}(y, \mu)  F(\mu)  F^+(\mu)  \over 
(2 \bar \Lambda  - \omega - i \epsilon) 
(2 \bar \Lambda^+ - \omega^\prime - i \epsilon) } \;, \label{pole}
\end{equation}
where $\mu$ is a renormalization scale and $F(\mu)$, $F^+(\mu)$ 
are the current-particle matrix elements
\begin{eqnarray} 
<0| J_5^v |B(v) > &=& F(\mu) \label{f} \\
<0| J_s^v |B_0(v) > &=& F^+(\mu) \label{fhat} \;\;\;. 
\eea
Notice that $F(\mu)$ is related to the $B$-meson leptonic decay 
constant $f_B$. 
The mass parameters 
$\bar \Lambda=M_B - m_b$ and $\bar \Lambda^+=M_{D_0} - m_c$ identify 
the position 
of the poles in $\omega$ and $\omega^\prime$, 
and can be interpreted as the binding energies of the $0^-$ and 
$0^+$ states.

$\Pi_\mu(\omega, \omega^\prime, y)$ can alternatively
 be computed in QCD in the 
Euclidean region in terms of 
perturbative and nonperturbative contributions
\be
\Pi = \Pi^{pert} + \Pi^{np} \;\;\;, \label{qcd}
\ee
where $\Pi^{np}$ represents the series of power corrections,
determined by quark and gluon vacuum condensates.
A sum rule for $\tau_{1/2}$ is obtained by matching the hadronic 
and the QCD representation of the correlator in a suitable 
range of Euclidean values of $\omega$ and $\omega^\prime$. 
By the same method, considering two-point correlators,
the constants $F(\mu)$ and $F^+(\mu)$ 
can be determined. 

Defining $\mu$-independent constants $F^+$ and $F$, we obtain:
$\bar \Lambda^+= 1.0 \pm 0.1$ GeV and 
$F^+ = 0.7 \pm 0.2$ GeV$^{3\over 2}$;
$\bar \Lambda = 0.5 \pm 0.1$ GeV and
$F = 0.45 \pm 0.05$ GeV$^{3\over 2}$.
The difference $\Delta= \bar \Lambda^+ - \bar \Lambda$
corresponds to the difference  
$m_{\bar D_0}-m_{\bar D}$, with
$\bar D$ and  $\bar D_0$ spin averaged states of the 
${1\over 2}^-$ and ${1\over 2}^+$ doublets. The central value
$\Delta=0.5$ GeV predicts $m_{\bar D_0}\simeq 2.45$ GeV 
with an uncertainty of about $0.15$ GeV.

Neglecting ${\cal O}(\alpha_s)$ corrections QCD sum rules  would give: 
$F^+=0.46 \pm 0.06$ GeV$^{3\over 2}$ \cite{cola92}
and $F^+=0.40 \pm 0.04$ GeV$^{3\over 2}$ and ${\bar \Lambda}^+=1.05 \pm 0.5$ 
GeV or ${\bar \Lambda}^+=0.90 \pm 0.10$ GeV \cite{dai}, 
by various choices of the interpolating currents.
$\alpha_s$ corrections are thus sizeable for the couplings 
(\ref{f}),(\ref{fhat}). The same is true for
$\Pi^{pert}$ in (\ref{qcd}).

Relativistic quark models 
give $F^+ \simeq 0.6 - 0.7$ GeV$^{3/2}$, whereas 
lower values are obtained: $F^+ \simeq 0.235$ GeV$^{3/2}$ 
if non relativistic models are employed
\cite{morenas}. 

The numerical result of the sum rule for $\tau_{1/2}$ 
including next-to-leading $\alpha_s$ corrections
is depicted in fig.(\ref{fig:tau12}); 
the shaded region essentially represents the 
theoretical uncertainty  of the calculation.
\begin{figure}[htb]
\vskip -1.0cm
\psfig{figure=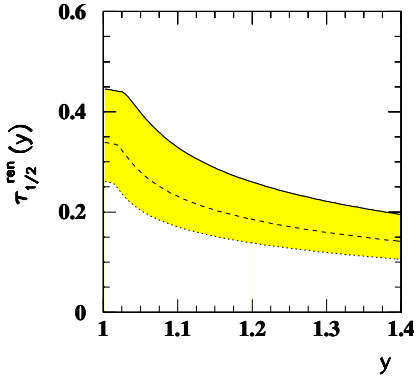,height=8.0 cm}
\vskip -2.5cm
\caption{Universal form factor $\tau_{1/2}(y)$}
\label{fig:tau12}
\end{figure}
Expanding the form factor near $y=1$ as 
\be
\tau_{1/2}^{ren}(y)=\tau_{1/2}(1) 
\Big(1-\rho^2_{1/2} (y-1)+c_{1/2} (y-1)^2\Big) \label{res}
\ee
we get: 
$\tau_{1/2}(1)=0.35\pm0.08$,
$\rho^2_{1/2}=2.5\pm 1.0$  and $c_{1/2}=3\pm 3$.

QCD sum rules at
${\cal O}(\alpha_s=0)$ \cite{cola92} gave the results
$\tau_{1/2}(1) \simeq 0.25$ and
$\rho^2_{1/2} \simeq 0.4$, which shows that $\alpha_s$ corrections,
although not negligible, do not dramatically affect the final result.

As for $\tau_{3/2}(y)$, a QCD sum rule analysis at the leading 
order in $\alpha_s$ \cite{cola92} gives
$\tau_{3/2}(1)\simeq 0.28$ and $\rho^2_{3/2}\simeq 0.9$.
Determinations  
of the universal form factors by constituent quark models
give results in a quite wide range:
$\tau_{1/2}(1)=0.06-0.40$, $\tau_{3/2}(1)=0.3-0.7$ \cite{morenas}.

Using $V_{cb}=3.9 \times 10^{-2}$ and 
$\tau(B)= 1.56$ ps, from (\ref{res}) we obtain
${\cal B}(B\to  D_0 \ell \bar \nu)=(5 \pm 3) \times 10^{-4}$ and
${\cal B}(B\to D^*_1 \ell \bar \nu)= (7 \pm 5) \times 10^{-4}$.

According to this result only a small fraction
of the inclusive semileptonic $B \to X_c$ 
decays  is represented by $B$ transitions into the
$s_\ell^P={1\over 2}^+$ charmed doublet. 
One cannot exclude that such processes might be in the reach of future
$B$-facilities \cite{babar}.
At present, data on semileptonic $B \to D^{**}$ decays
exist only for the $s_\ell^P={3\over 2}^+$ doublet,
since the $s_\ell^P={1\over 2}^+$ doublet
is not distinguished from the non-resonant charmed background
due to the large width.

We conclude by observing that predictions derived within HQET
must always be supported by the calculation of $1/m_Q$ as well as radiative 
corrections. The role of both depend on the specific situation one is 
faced with. For the 
form factor $\tau_{1/2}(y)$, using QCD sum rules in the framework 
of HQET, we have obtained that,
similar to the case of the Isgur-Wise function, radiative 
corrections are quite under control while they affect 
considerably  the value of the leptonic constants.


\begin{thebibliography}{9}

\bibitem{review}
For reviews see
M. Neubert, Proceedings of the 34th Course of the Erice International School
of Subnuclear Physics, Erice 1996, page 98;
M. Wise, hep-ph/9805468;
R. Casalbuoni et al., Phys. Rep. {\bf 281} (1997) 145. 
 
\bibitem{pdg}
Review of Particle Properties, 
C. Caso et al., Eur. Phys. J. C {\bf 3} (1998) 1.

\bibitem{falk92}
A.F. Falk et al., Phys. Lett. B {\bf 292} (1992) 119.

\bibitem{colangelo95}
P. Colangelo {\it et al.}, Phys. Rev. D {\bf 52} (1995)  6422.
V. M. Belyaev et al., Phys. Rev. D {\bf 51} (1995) 6177.
P. Colangelo and F. De Fazio, Eur. Phys. J. C {\bf 4} (1998) 503.

\bibitem{isgur91}
N. Isgur and M. Wise, Phys. Rev. D {\bf 43} (1991) 819.

\bibitem{colangelo98}
P. Colangelo, F. De Fazio and N. Paver, hep-ph/9804377,
to appear on Phys. Rev. D.

\bibitem{neubert93}
M. Neubert, Phys. Rev. D {\bf 47} (1993) 4063. 

\bibitem{cola92}
P. Colangelo, G. Nardulli and N. Paver, Phys. Lett. B {\bf293} (1992) 207.
P. Colangelo et al., 
Phys. Lett. B {\bf 269} (1991) 204.

\bibitem{dai}
Y.B. Dai et al., Phys. Lett. B  {\bf 390} (1997) 350; 
Phys. Rev.  D {\bf 55} (1997) 5719;  hep-ph/9705223.

\bibitem{morenas}
V.Morenas et al., Phys. Rev. D {\bf 56} (1997) 5668, hep-ph/9710298.
A. Deandrea et al., Phys. Rev. D {\bf 58} (1998) 34004.

\bibitem{babar}
The BaBar Physics Book, SLAC-R-504.

\end{thebibliography}
\end{document}